\documentstyle[12pt]{article}
\begin{document}

\title {PRECISION TESTS OF THE MSSM.
\thanks{Plenary talk given by S. Pokorski at the ``Beyond the Standard
        Model IV'', Lake Tahoe, CA, December 1994.}
\thanks{Supported in part by the Polish Committee for Scientific
        Research and European Union Under contract CHRX-CT92-0004.}}
\author{Piotr H. Chankowski \\
Institute of Theoretical Physics, Warsaw University\\
ul. Ho\.za 69, 00--681 Warsaw, Poland.\\
\\
Stefan Pokorski
\thanks{On leave of absence from
Institute of Theoretical Physics, Warsaw University}
\\
Max--Planck--Institute f\"ur Physik\\
Werner -- Heisenberg -- Institute \\
F\"ohringer Ring 6,
80805 Munich, Germany
}

\maketitle

\vspace{-12cm}
\begin{flushright}
{\bf IFT--95/5}\\
{\bf March 1995}
\end{flushright}
\vspace{12cm}

\begin{abstract}
We present the results of a first global fit to the electroweak
observables in the MSSM. The best fit selects either very low or
very large values of ~$\tan\beta$ ~and, correspondingly, chargino
(higgsino--like) and stop or the ~$CP-$odd Higgs boson are within the
reach of LEP 2. Moreover, the best fit gives ~
$\alpha_s(M_Z)=0.118^{+0.005}_{-0.010}$, ~which is lower than the
one obtained from the SM fits. The overall fit is excellent ~
($\chi^2=7.2$ ~for 15 d.o.f. as compared to ~$\chi^2=11$ ~in the SM).
Those results follow from the fact that in the MSSM one can increase
the value of ~
$R_b\equiv\Gamma_{Z^0\rightarrow\bar bb}/\Gamma_{Z^0\rightarrow hadrons}$ ~
{\it without} modyfying the SM predictions for other observables.
\end{abstract}

\newpage
Precision tests of the MSSM have been discussed by several groups
\cite{ALBA,LANLUO,ABC,ELLIS,CW}. Here we present for the first
time the results of a global fit to the electroweak observables.

We discuss the MSSM as an effective low energy theory,
irrespectively of its high energy roots, and do not use any GUT scale
boundary conditions to constrain the parameters of the MSSM. On the
contrary, we believe the set of low energy parameters suggested by the
precision data may give an interesting hint on physics at the GUT scale.

Our strategy is analogous to the one often used for the SM:
in terms  of the best measured observables ~$G_F$, ~
$\alpha_{EM}$, ~$M_Z$ ~and the  less well known ~
$m_t$, ~$\alpha_s(M_Z)$ ~and the large number of additional free parameters
in the MSSM  such as ~$\tan\beta$, ~$M_A$, ~soft SUSY breaking scalar masses,
trilinear couplings etc. we calculate in the MSSM the observables ~$M_W$, ~
all partial widths of ~$Z^0$ ~and all asymetries at the ~$Z^0$ ~pole. This
calculation is performed in the on--shell renormalization scheme
\cite{HOL,HOLL,MY_H} and with
the same precision as the analogous calculation in the SM, i.e. we include
all supersymmetric oblique and process dependent one--loop corrections
and also the leading higher order effects.

Similar programme has often been discussed in the context of the SM
\cite{ALBA,LANER,EWG} with the parameters ~$m_t$, ~$\alpha_s(M_Z)$ ~and
$M_h$ ~(or some of them) to be determined by a fit to the data.
Let us first review the results in the SM with the emphasis on those
features which are relevant for the supersymmetric extension.

The experimental
input (i.e. experimental values for the electroweak observables,
their errors and correlation matrices) used in the fits is summarized
in ref. \cite{EWG}. For the top quark
mass we use the CDF result ~$m_t=(176\pm13)$ GeV \cite{CDF_MT}.
In the most recent
data of ref. \cite{EWG,MORIOND_95} there are two significant changes
compared to the data of the Glasgow conference \cite{GLA,MW}: the central
value of ~$M_W$ ~and of ~$A_{FB}^{0~b}$ ~have increased and they now
read ~$M_W=(80.33\pm0.17)$ GeV, ~$A_{FB}^{0~b}=0.1015\pm0.0036$. ~

In Table 1 we present the results of our global fit without and with
the SLD result for the electron left--right asymetry \cite{SLD} included.
In both cases the lower limits on ~$M_h$ ~come from the unsuccesful direct
Higgs boson search.

\begin{center}
{\bf Table 1.} Results of the fits to the data \cite{EWG}.
All masses in GeV.
\vskip 0.2cm
\begin{tabular}{||l||l|l|l|l|l|l|l|l||} \hline
 fit             &
$m_t$      &
$\Delta m_t$   &
$M_h$           &
$\Delta M_h$    &
$\alpha_s(M_Z)$ &
$\Delta\alpha_s(M_Z)$ &
$\chi^2$         &
d.o.f            \\ \hline
$-$ SLD  & 168  & 12 & 121 & $^{+207}_{-58}$ & 0.124& 0.005 & 11.1& 15\\ \hline
$+$ SLD  & 166  & 11 &  63 & $^{+97}_{-0}$  & 0.123& 0.005 & 16.5 &16 \\ \hline
\end{tabular}
\end{center}

In the following, in fits in the MSSM we do not include
the SLD result.

The fitted value for ~$M_h$ ~ results in a very transparent
way from a combination of effects which can be organized into the
following two--step description \cite{MY_SMH}.
A fit to the measured ~$M_W =
(80.33\pm0.17)$ GeV ~\cite{MORIOND_95} and to all measured electroweak
observables but ~
$R_b\equiv\Gamma_{Z^0\rightarrow\bar bb}/\Gamma_{Z^0\rightarrow hadrons}$ ~
gives ~$\chi^2$ ~values which are almost
{\it independent} of the value of the top quark mass ~$m_t$ ~
in the broad range (150--200 GeV) and with the best value of ~
$\log M_h$ ~which is almost linearly correlated with ~$m_t$. ~
This is shown in Fig.1.
The most recent experimental results \cite{EWG,MORIOND_95}, in particular
the increase in the  central value of ~$M_W$ ~by ~100 MeV ~and the new
measurement of ~$A_{FB}^{0~b}$, ~evolved in the direction of requiring
smaller ~$M_h$ ~for a given, fixed ~$m_t$ ~(or larger ~$m_t$ ~for a given,
fixed ~$M_h$). ~

The ~$(m_t, M_h)$ ~correlation is the most solid result of the fits
which does not depend on whether ~$R_b$ ~and/or ~$m_t$ ~measurement of
the CDF \cite{CDF_MT} are included into the fits. It points toward
relatively light Higgs boson for ~$m_t$ ~in the range ~$(170 - 180)$ GeV. ~

The ~$\chi^2$ ~is {\it flat} as a function of ~$m_t$ ~ unless ~$R_b$ ~
and/or the directly measured in Fermilab mass of the top quark are
included into the fit. These are the only two measured observables
which in the present case introduce visible ~$\chi^2$ ~dependence
as a function of ~$m_t$ ~
and can, therefore, put indirectly (by constraining ~$m_t$) ~relevant
overall limit on the Higgs mass ~$M_h$.
This is also  clearly seen in Fig.1.

Finally we can interpret the SM fits as the MSSM fits with all
superpartners heavy enough to be decoupled. Supersymmetry then just
provides a rationale for a light Higgs boson: $M_h \sim {\cal O}$(100 GeV).
We can conclude that the MSSM with heavy
enough superpartners is expected to give a perfect fit to the precision
electroweak data, with $m_t\sim170$ GeV ~(depending slightly on the value
of ~$\tan\beta$) ~and moreover $m_t<$185 GeV at 95 \% C.L.

This is seen in Fig.2 where we show the ~$\chi^2$ ~values in the MSSM
with the proper dependence of ~$M_h$ ~on ~$m_t$, ~$\tan\beta$ ~and
SUSY parameters included \cite{ZWIR}, for a fit of ~$m_t$ ~and ~
$\alpha_s(M_Z)$ ~with all SUSY mass parameters fixed at 1.5 TeV.
The ~$\chi^2$ ~values in the minima are exactly the same as in the SM
fit !

It is clear from this comparison, that the only room for
improvement in the MSSM compared to the SM is in the value of ~$R_b$.
We can ask the following two questions:

\noindent a) can we improve ~$R_b$ ~without destroying the excellent
fit to the other observables?

\noindent b) if we achieve this goal, what are the limits on
sparticle masses?

Moreover it is well known that additional contributions to ~
$\Gamma_{Z^0\rightarrow\bar bb}$ ~would lower the fitted value of ~
$\alpha_s(M_Z)$ ~\cite{BV,SHIF}, in better agreement with its
determination from low energy data \cite{LOW}.

We begin with a brief overview of the SUSY corrections to the electroweak
observables. Although
the MSSM contains many free SUSY parameters, there are remarkable regularities
in SUSY corrections. Following our strategy
of calculating all electroweak observables in terms of ~$G_F$, ~$M_Z$ ~and ~
$\alpha_{EM}(0)$ ~one can establish the following ``theorems'' for the
predictions in the MSSM:

\vskip 0.2cm
\noindent 1.) $(M_W)^{MSSM} \geq (M_W)^{SM}$.
As explained in ref.\cite{MY_DR}, its origin lies in additional sources of
the custodial $SU_V(2)$ violation in the squark and slepton sectors:
\begin{eqnarray}
M^2_{\tilde l_L} = M^2_{\tilde\nu} + t_{\beta} M^2_W, ~~~~~
M^2_{\tilde t_L} = M^2_{\tilde b_L} + m^2_t - m^2_b - t_{\beta}M_W^2
\label{eqn:m_split}
\end{eqnarray}
where $t_{\beta}\equiv(\tan^2\beta-1)/(\tan^2\beta+1)$, which contribute to
$\Delta\rho$ with the same sign as the $t - b$ mass splitting.
It should be stressed that the
supersymmetric prediction for $M_W$ is merely sensitive
to $M_{\tilde l_L}$ and $M_{\tilde q_L}$ -- masses of the left--handed
sleptons and third generation left--handed squarks. The dependence
on the right--handed sfermion masses
enters only through the left--right mixing. This also means that the
predicted $M_W$ is almost insensitive to the masses of squarks of the first
two generations: in their left--handed components there is no source of large
$SU_V(2)$ violation. Also, the predictions for $M_W$ are rather weakly
dependent on the chargino and neutralino masses $m_{C^\pm}$, $m_{N^0}$
and the Higgs sector parameters
\footnote{This is due to generically weak $SU_V(2)$ breaking effects in
          these sectors.}.
\vskip 0.2cm
\noindent 2.) Another effect of supersymmetric corrections is that ~
$(\sin^2\theta^{\it l})^{MSSM} \leq (\sin^2\theta^{\it l})^{SM}$ where
$\sin^2\theta^{\it l}$ ~can be determined from the
on--resonance forward--backward asymmetries
\begin{eqnarray}
A_{FB}^{0~\it l} = {3\over4}{\cal A}_e{\cal A}_{\it l}
{}~~{\rm where} ~~
{\cal  A}_f = {2x_f\over 1+ x^2_f}
\end{eqnarray}
with ~$x_f = 1 - 4|Q_f|\sin^2\theta^f$.
In general, in the on--shell renormalization scheme and with the loop
corrections included we get:
\begin{eqnarray}
\sin^2\theta^{\it l}=
\left(1 - {M^2_W\over M^2_Z}\right)\kappa_{UN}(1+ \Delta\kappa_{NON})
\label{eqn:sin_for}
\end{eqnarray}
where $\kappa_{UN}$ contains universal ``oblique'' corrections and
$\Delta\kappa_{NON}$ -- genuine (nonuniversal) vertex corrections which
are in this
case negligibly small. By using explicit form of $\kappa_{UN}$ ~one can derive
the following relation:
\begin{eqnarray}
(\sin^2\theta^{\it l})^{MSSM} = (\sin^2\theta^{\it l})^{SM}\times
\left[1 - {c^2_W\over c^2_W - s^2_W}
\left(\Delta\rho\right)^{SUSY} + ...\right]
\end{eqnarray}
We see therefore, that the supersymmetric predictions for ~
$\sin^2\theta^{\it l}$ ~are correlated with the predictions for ~$M_W$ ~
through the value of ~$\Delta\rho$ ~
and they are sensitive to the same supersymmetric parameters.
\vskip 0.2cm

3.) Similarly, the asymmetries in the quark channel are given by the product
\begin{eqnarray}
A_{FB}^{0~\it q} = {3\over4}{\cal A}_e{\cal A}_q
\end{eqnarray}
If ~
$(\sin^2\theta^{\it l})^{MSSM}=(\sin^2\theta^{\it l})^{SM} - \varepsilon$ ~
and ~$(\sin^2\theta^b)^{MSSM}=(\sin^2\theta^b)^{SM} - \delta$ ~
then it is easy to show that
\begin{eqnarray}
(A_{FB}^{0~\it b})^{MSSM} =(A_{FB}^{0~\it b})^{SM} \times
\left(1 + {\varepsilon\over1-4\sin^2\theta^{\it l}} + 0.2\delta\right)
\end{eqnarray}
Thus, supersymmetric corrections to ~$A_{FB}^{0~\it b}$ ~are essentially
determined
by the corrections to ~$\sin^2\theta^{\it l}$ ~and give
the third ``theorem'': ~
$(A_{FB}^{0~\it b})^{MSSM}\geq (A_{FB}^{0~\it b})^{SM}$. ~

At this point it is important to observe that the trends in the MSSM
expressed by the above three theorems can only make the comparison of the
MSSM predictions with the data worse than in the SM (as for ~$m_t > 160$
GeV ~they go against
the trend of the data!). Thus we can expect to get relevant lower limits
on the left--handed squark and slepton masses, which are the parameters
most relevant for the observables ~$M_W$, ~$\sin^2\theta^{\it l}$ ~and ~
$A_{FB}^{0~\it b}$ ~(of course, for large enough masses, we recover the SM
predictions). These limits are amplified by the dependence of ~$\Gamma_Z$ ~
on ~$M_{\tilde q_L}$ ~and ~$M_{\tilde {\it l}_L}$. ~As discussed above, the
sensitivity of the each one of those  observables to the other parameters
is weaker but accumulates itself and becomes non-negligible in the global
fit. The most important and interesting is the dependence on the chargino
mass and its composition. One can see that a light higgsino (and only
higgsino) does not worsen the predictions for ~
$M_W$, ~$\sin^2\theta^{\it l}$, ~$A_{FB}^{0~\it b}$ ~and, for a heavy top
quark and light Higgs boson it can significantly improve the fit to ~
$\Gamma_Z$ ~\cite{BARB} (this is a ~$Z^0$-wave function renormalization
effect which acts similarly to a heavier Higgs boson for heavier $t$ quark
i.e. its contribution makes ~$\Gamma_Z$ ~smaller). ~

Finally, let us discuss the
corrections to the ~$Z^0\rightarrow\overline b b$ ~
vertex which contribute to the observables ~$R_b$ ~and ~$\sin^2\theta^b$. ~
The latter can be extracted e.g. from ~$A_{FB}^{0~\it b}$ ~once ~
$\sin^2\theta^{\it l}$ ~is known or, more conveniently, from
the measurements of the the polarized  forward--backward asymmetry ~
$A_{FB}^{pol ~b} = {\cal A}_b$ ~at SLAC. The vertex corrections
contribute to ~
$\Delta\kappa_{NON}$ ~in eq. (\ref{eqn:sin_for}) and can dramatically
change the result for ~$\sin^2\theta^b$ ~compared to its value predicted
in the SM and can give relevant contribution to ~$R_b$.

In the MSSM there are three types of important corrections to the vertex ~
$Z^0\overline b b$ ~\cite{BF}:

\noindent a) charged and neutral Higgs boson exchange; for low ~
$\tan\beta$ ~
and light ~$CP-$odd Higgs boson ~$A^0$ ~this contribution is negative (the ~
$\Gamma_{Z\rightarrow\bar bb}$ ~is decreasing below its SM value) whereas
for very light ~$A^0$ ~(50 -- 80 GeV) and very large values of the ~
$\tan\beta$ ~
($\sim50$) ~the interplay of charged and neutral Higgs bosons is strongly
positive;

\noindent b) chargino -- stop loops; for heavy top quark and low ~
$\tan\beta$ ~
they can contribute significantly (and positively) for light chargino (if
higgsino-like) and light right-handed top squark (this follows from the
Yukawa chargino--stop--bottom coupling); in the case of large $\tan\beta$
this contribution is smaller than for small ~$\tan\beta$ ~but the total
contribution to the ~$Z^0\bar b b$ ~vertex can be amplified by

\noindent c) neutralino -- sbottom (if light) loops.

Thus, in the MSSM the value of ~$R_b$ ~can in principle be significantly
larger than in the SM for very low or very large values of ~$\tan\beta$, ~
light (higgsino-like) chargino, and ~$\tilde t_R$ ~and/or very light ~
$A^0$ ~(for large ~$\tan\beta$) ~\cite{KKW}. It is insensitive to ~
$M_{\tilde q_L}$ ~and ~$M_{\tilde {\it l}_L}$.

In summary, in MSSM the electroweak observables exhibit
certain ``decoupling'': all of them but ~$R_b$ ~are sensitive mainly
to the left--handed slepton and third generation squark masses and
depend weakly on the right--handed squarkand slepton masses, gaugino
and Higgs sectors; on the contrary, ~$R_b$ ~depends strongly just
on the latter set of variables and very weakly on the former. We can
then indeed expect to increase the value of ~$R_b$ ~ without destroying
the perfect fit of the SM to the other observables. However,
chargino, right--handed stop and charged Higgs boson masses also
are crucial variables for the decay ~$b\rightarrow s\gamma$ ~ and this
constraint has to be included \cite{BSG_BG}. In addition,
in the parameters space
which gives light higgsino--like charginos also neutralinos are
higgsino--like and the contribution of ~$Z^0\rightarrow N_i^0N_j^0$ ~
to the total ~$Z^0$ ~width must be included (we also impose the
constraint that ~$N^0_1$ ~is the LSP).

For large ~$\tan\beta$ ~the value of ~$R_b$ ~can be enhanced with
very light ~
$C^\pm$, ~$\tilde t_R$ ~and/or ~$M_A\sim 50 - 70$ GeV. ~The chargino--stop
loop contribution to ~$BR(b\rightarrow s\gamma)$ ~reads:
\begin{eqnarray}
A_{C^\pm}\sim{1\over2}\tan\beta\left({m_t\over M_{\tilde t_1}}\right)^2
\left({A_t\over\mu}\right)\log{M_{\tilde t_1}^2\over\mu^2}
\end{eqnarray}
and, unless ~$A_t\sim0$, ~it is very large for ~$m_C$, ~
$M_{\tilde t_R} <{\cal O}(100$GeV). ~
For a light ~$A^0$, ~the charged Higgs boson contribution
to  ~$BR(b\rightarrow s\gamma)$ ~is also too large \cite{BMMP}
and has to be cancelled
by the  chargino--stop loop contribution. Note however,
that given the order of magnitude of the latter, this cancellation may
occur for rather heavy chargino and/or ~$\tilde t_R$ ~and therefore small ~
$R_b$, ~or, again, for ~$A_t <{\cal O}(M_Z)$. ~

Thus, one obtains acceptable ~$BR(b\rightarrow s\gamma)$ ~due to
cancellations between ~$W^{\pm}$, ~$H^{\pm}$ ~and ~$C^{\pm}$ ~and ~
$\tilde t_R$ ~loops. The net impact on the allowed parameters space depends
quite strongly on the values of ~$m_t$ ~and ~$\alpha_s(M_Z)$. ~
Generically, the values of  ~$BR(b\rightarrow s\gamma)$ ~are in the lower
edge of the allowed range.

Let us present some quantitative results.
We have fitted the value of ~$\alpha_s$, ~$\tan\beta$, ~$m_t$ ~and SUSY
parameters. The first observation is that the ~$\chi^2$ ~is generically
always better than in the SM and this is due to the
improvement in ~ $R_b$. ~
The dependence of the ~$\chi^2$ ~on ~$\tan\beta$ ~for several values of ~
$m_t$ ~(and scanned over the other parameters) is shown in Fig.3. The best
fit is obtained in two regions of very low (close to the  quasi--IR
fixed point for a given top quark mass) and very large ~$(\sim m_t/m_b)$ ~
$\tan\beta$ ~values. We get, respectively ~
$m_t = 170^{+10}_{-5}$ GeV ~and ~$m_t = (165\pm10)$ GeV. ~
The values of ~$R_b$ are in the range ~0.218--0.219. ~

The global dependence of the ~$\chi^2$ ~on ~$m_t$ ~ is shown in Fig. 4b
and gives  ~$m_t = 166^{+12}_{-9}$ GeV. ~For the best fit in the low ~
$\tan\beta$ ~region we obtain strong bounds on chargino and the lightest
stop masses. They are shown in Fig.5. In the large ~$\tan\beta$ ~region
the most relevant variable is the pseudoscalar Higgs boson mass ~$M_A$ ~
and we get ~$M_A < 70$ GeV ~at ~$1\sigma$ ~level in the other analyses
\cite{SOLA}.

Finally, an important issue is the obtained value of ~$\alpha_s(M_Z)$. ~
It is well known \cite{BV,SHIF} that additionaal contributions to ~
$\Gamma_{Z\rightarrow\bar bb}$ ~should lower the fitted value of ~
$\alpha_s(M_Z)$. ~This effect is indeed observed in our fits and brings
the obtained value of ~$\alpha_s(M_Z)$ ~closer to the values
obtained from low energy data \cite{LOW}.
In Fig. 6a we show ~$\chi^2$ ~values as
a function of ~$\alpha_s(M_Z)$ ~for different values of ~$\tan\beta$ ~
and in Fig. 6b the global dependence of ~$\chi^2$ ~on  ~$\alpha_s(M_Z)$. ~
For low and intermediate ~$\tan\beta$ ~values we get  ~$\alpha_s(M_Z)=
0.118\pm0.005$ ~and for ~$\tan\beta\approx m_t/m_b$, ~$\alpha_s(M_Z)=
0.113\pm0.008$. ~

Finally in Fig.7 we show the lower ~1 ~and ~2$\sigma$ ~limits on
left--handed sbottom and left--handed slepton masses for different ~
$m_t$ ~and ~$\tan\beta$. ~

In summary, a MSSM fit to the electroweak observables is excellent ~
($\chi^2 \sim 7.2$ ~for 15 d.o.f.; in the SM we have \cite{MY_SMH} ~$\chi^2=
11$). ~This is due to higher than in the SM values of ~$R_b$ ~obtained
in the MSSM, without destroying the agreement in the other observables.
The best fit selects very particular regions of the parameters space:
either very low or very large values of ~$\tan\beta$ ~and correspondingly,
chargino (higgsino--like) and stop or the ~$CP-$odd ~Higgs boson masses
are within the reach of LEP 2. Moreover the best fit gives ~$\alpha_s(M_Z)=
0.118^{+0.005}_{-0.010}$, ~a value which
is lower than the one obtained from
the SM fits and in agreement with the low energy data. Low values
of ~$\alpha_s(M_Z)$ ~is correlated with the presence of the additional
contribution to the ~$\Gamma_{Z\rightarrow\bar bb}$ ~\cite{BV,SHIF}.
\vskip 2.0cm

{\bf Acknowledgments:} P.Ch. would like to thank
 Max--Planck--Institut f\"ur
Physik for warm hospitality during his stay in Munich where
part of this work was done.

\newpage

\noindent {\bf FIGURE CAPTIONS}
\vskip 0.5cm

\noindent {\bf Figure 1.}

\noindent Limits in the ~$(m_t, M_h)$ ~plane in the SM.
Unclosed lines show the ~1$\sigma$ ~limits from the fit
without the ~$R_b$ ~and ~$m_t$ ~measurements included.
Ellipses show the ~1$\sigma$ ~and ~2$\sigma$ ~limits from
the fit with the ~$R_b$ ~and ~$m_t$ ~measurements included.
\vskip 0.3cm

\noindent {\bf Figure 2.}

\noindent ~$\chi^2$ ~as a function of ~$m_t$ ~
for different values of ~$\tan\beta$ ~in the MSSM with
very heavy superparticles.
Dashed lines show  ~$\chi^2$ ~without the ~$R_b$ ~measurement
in the fit.
\vskip 0.3cm

\noindent {\bf Figure 3.}

\noindent Dependence of ~$\chi^2$ ~on ~$\tan\beta$ ~for different
values of ~$m_t$.
\vskip 0.3cm

\noindent {\bf Figure 4.}

\noindent Dependence of ~$\chi^2$ ~on ~$m_t$: ~a) for different
values of ~$\alpha_s(M_Z)$, ~b) for best ~$\alpha_s(M_Z)$ ~for
each ~$m_t$.
\vskip 0.3cm

\noindent {\bf Figure 5.}

\noindent 1 ~and ~2$\sigma$ ~limits on chargino and lighter stop
masses for ~$m_t=180$ GeV, ~$\tan\beta=1.6$ (IR), ~and ~
$\alpha_s(M_Z)=0.12$.
\vskip 0.3cm

\noindent {\bf Figure 6.}

\noindent a) Dependence of ~$\chi^2$ ~on ~$\alpha_s(M_Z)$ ~
for different values of ~$\tan\beta$, ~b)  global dependence
of ~$\chi^2$ ~on ~$\alpha_s(M_Z)$. ~
\vskip 0.3cm

\noindent {\bf Figure 7.}

\noindent 1 ~and ~2$\sigma$ ~limits on left--handed sbottom and
slepton masses for different values of ~$m_t$ ~and ~$\tan\beta$.
\vskip 0.3cm

\end{document}